\documentclass[10pt,a4paper]{article}
\usepackage[utf8]{inputenc}
\usepackage[english]{babel}
\usepackage{amsmath}
\usepackage{amsfonts}
\usepackage{amssymb}
\usepackage{graphicx}
\usepackage[left=2cm,right=2cm,top=2cm,bottom=2cm]{geometry}
\usepackage{authblk}
\usepackage{graphicx}
\usepackage{subcaption}
\graphicspath{ {./Images_paper/} }
\title{\bf{On the use of PN Ranging with High-rate Spectrally-efficient Modulations}}
\usepackage{fancyhdr}
\usepackage{lastpage}

\author[a,*]{{ B. Ripani}}
\author[b]{{ A. Modenini}}
\author[a]{{ R. Garello}}
\author[a]{{ G. Maiolini Capez}}
\author[a]{{ G. Montorsi}}
\affil[a]{\textit{Politecnico di Torino, Corso Duca degli Abruzzi 24, Torino, Italy}\newline {\{barbara.ripani, roberto.garello, gabriel.maiolinicapez, guido.montorsi\}@polito.it}}
\affil[b]{\textit{European Space Agency, Keplerlaan 1, Noordwijk, The Netherlands}\newline andrea.modenini@esa.int}
\affil[*]{Corresponding Author}

\date{}
\begin{document}

	\maketitle
	\thispagestyle{fancy}
	
	\begin{abstract}
		In this paper, we study the feasibility of coupling the PN ranging with filtered high-order modulations, and investigate the simultaneous demodulation of a high-rate telemetry stream while tracking the PN ranging sequence. Accordingly, we design a receiver scheme that is able to perform a parallel cancellation, in closed-loop, of the ranging and the telemetry signal reciprocally. 
From our analysis, we find that the non-constant envelope property of the modulation causes an additional jitter on the PN ranging timing estimation that, on the other hand, can be limited by properly sizing the receiver loop bandwidth.

Our study proves that the use of filtered high-order modulations combined with PN ranging outperforms the state-of-the-art in terms of spectral efficiency and achievable data rate, while having comparable ranging performance.
\newline

		
	\end{abstract}

\section{Introduction}
	Space agencies are currently replacing transparent ranging systems with regenerative schemes to meet future Space Research (SR) mission needs. For this purpose, the Consultative Committee for Space Data Systems (CCSDS) defined the Pseudo-Noise (PN,~ \cite{CCSDSPNB,CCSDSPNG}) ranging standard, a state-of-the-art positioning technique. This standard, already implemented in the ESA missions BepiColombo and Solar Orbiter, demonstrated an accuracy as high as a few centimetres during its validation~\cite{cappuccio2019first,8895263}. Additionally, the possibility of coupling the PN ranging with a Gaussian Minimum Shift Keying (GMSK,~\cite{Murota1981}) modulated signal allows simultaneous reception of high-rate telemetry streams and orbit determination.

However, Near Earth SR missions transmitting telemetry in X-Band, in agreement with frequency assignment regulations~\cite{SFCG}, are constrained to 10 MHz of bandwidth.
Thus, when using GMSK, these missions are limited to a data rate of 8.7 Mbps, penalizing payload data generation (see the case of GAIA mission~\cite{GAIA}). To exceed this bound, it is clear that high-order modulation schemes must be adopted.  However, their compatibility with PN ranging (to the best of the authors' knowledge) was never investigated.

In this paper, we study the coupling of PN ranging with filtered (amplitude) Phase-Shift Keying (PSK/APSK) signals, by taking the CCSDS standard 131.2~\cite{CCSDSSCCC_G,CCSDSSCCC} as a reference. Namely, we define a receiver scheme that is able to simultaneously demodulate the PSK/APSK stream, with modulation orders as high as 64, and track the PN ranging sequence. This is done by performing a parallel closed-loop cancellation of the PN ranging sequence over the telemetry symbols and vice-versa.
To evaluate the performance of the system we designed, we assess both the Bit Error Rate (BER) and the ranging timing jitter, revealing the presence of a noise floor due to the non-constant envelope of PSK/APSK modulations. By modeling this phenomenon we derive a closed-form expression that constitutes an upper-bound for the ranging jitter.

In this work, we find the performance of the proposed scheme to be comparable to the classical GMSK approach, while enabling much higher data rates, thus paving the way towards a new generation of SR missions with more ambitious scientific objectives.

\section{System Model}
\begin{figure}
	\centering
		\includegraphics[width=0.6\columnwidth]{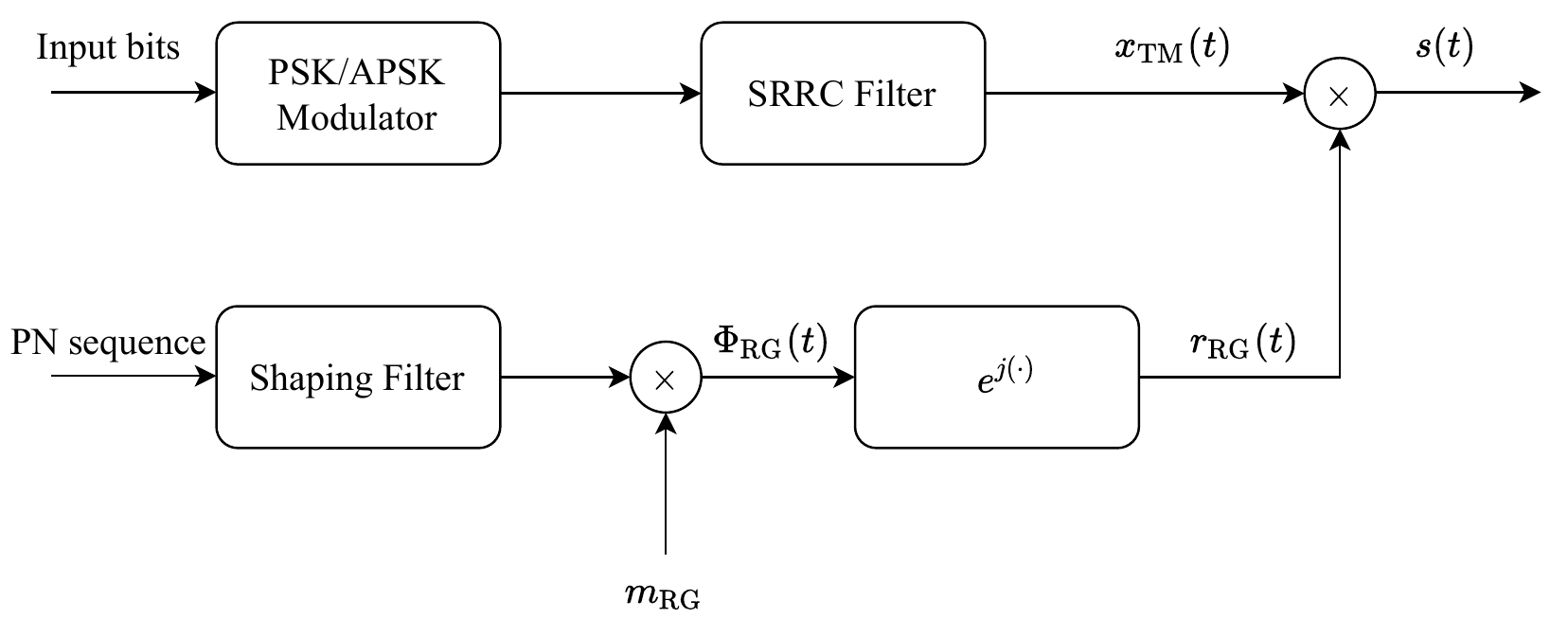}
		\caption{Transmitter block diagram.}
		\label{fig:Tx}
	\end{figure}

We consider the transmission system described by the block diagram reported in Figure~\ref{fig:Tx}. The complex base-band modulated telemetry signal, with unitary power, is defined as
\begin{equation}
	x_{\textrm{TM}}(t) =  \sum_k{a_k p(t- kT)} \,,
\end{equation}
where $p(t)$ is the Square Root Raised Cosine (SRRC)  shaping pulse, and $\{a_k\}$ the sequence of telemetry symbols, each transmitted at symbol time $k T$, and belonging to a complex PSK/APSK constellation. Instead, the phase-modulated ranging signal $r_{\textrm{RG}}(t)$ is defined as $
r_{\textrm{RG}}(t) = e^{j \Phi_{\textrm{RG}}(t)} $ and its phase can be expressed as  
\begin{equation}
 \Phi_{\textrm{RG}}(t) = m_{\textrm{RG}} \sum_k{c_k h(t-kT_c)} \,,
 \label{eq:phaseRG}
\end{equation}
where $m_{\textrm{RG}}$ is the ranging modulation index, $h(t)$ the sinusoidal chip shaping pulse defined in~\cite{CCSDSPNB}, and $\{c_k\}$ the sequence of  chips transmitted at chip time $k T_c$. Note that, in line with CCSDS recommendation 2.4.22A~\cite{CCSDSRF}, we imposed $T_c \neq n T$, with $n \in \mathbb{Z}^{+}$.
 
Therefore, 
by considering a transmission over the Additive White Gaussian Noise (AWGN) channel, the received signal reads
\begin{equation}
y(t) = x_{\textrm{TM}}(t) \cdot r_{\textrm{RG}}(t - \tau_{\textrm{RG}})  + w(t) \,,
\end{equation}
where $w(t)$ is white Gaussian noise with spectral density equal to $N_0$ and $\tau_{\textrm{RG}} \in [-\frac{T_c}{2},\frac{T_c}{2})$ is an arbitrary ranging timing delay.

\section{Receiver for PSK/APSK-modulated signals and PN Ranging} \label{sec:PSK/APSK+ranging}
\begin{figure}
	\centering
		\includegraphics[width=0.6\columnwidth]{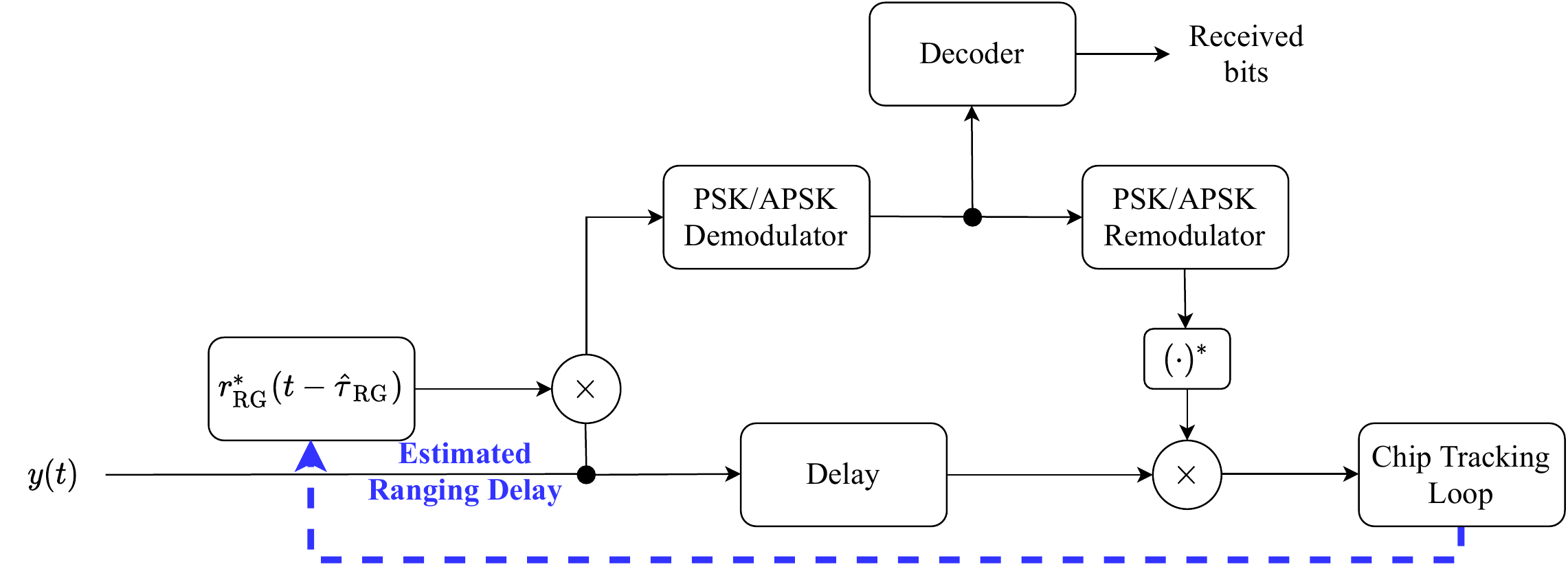}
		\caption{Receiver block diagram.}
	\label{fig:Rx}
\end{figure}

	We implement a modified version of the receiver defined in \cite{CCSDSGMSKPN_G} that performs parallel closed-loop cancellation of telemetry and ranging signals from the received stream. The receiver's block diagram is shown in Figure~\ref{fig:Rx}. 
	
	The ranging component is removed from the received signal by using a locally generated replica of the ranging signal.
	Mathematically,

\begin{equation}
	y(t)r^*_{\textrm{RG}}(t - \hat{\tau}_{\textrm{RG}}) = x_{\textrm{TM}}(t) + w(t)r^*_{\textrm{RG}}(t-\hat{\tau}_{\textrm{RG}}) \,, \label{eq:PNran_canc}
\end{equation}

being $r^*_{\textrm{RG}}(t)$ the ranging signal's complex conjugate, locally generated using the estimated time delay $\hat{\tau}_{\textrm{RG}}$. It is easy to see that, being $r_{\textrm{RG}}(t)$ a phasor, the resultant process $w(t)r^*_{\textrm{RG}}(t-\hat{\tau}_{\textrm{RG}})$ is still white Gaussian, and $x_{\textrm{TM}}(t)$ is perfectly recovered as long as $\hat{\tau}_{\textrm{RG}}=\tau_{\textrm{RG}}$,  since $|r_{\textrm{RG}}(t-\tau_{\textrm{RG}})|^2 = 1$.
The telemetry stream thus obtained is demodulated to extract data, and then re-modulated to obtain the sequence of symbols. With it, we generate the complex conjugate $x^*_{\textrm{TM}}(t)$, necessary to sub-optimally cancel the telemetry component from the received signal. We can write
\begin{equation} 
y(t)x^*_{\textrm{TM}}(t) =  r_{\textrm{RG}}(t-\tau_{\textrm{RG}})|x_{\textrm{TM}}(t)|^2 + \mathcal{W}(t) \,, \label{eq:PNtm_canc}
\end{equation}
where $\mathcal{W}(t) = x^*_{\textrm{TM}}(t)w(t)$ can be proved\footnote{It is pointed out that the signal $x_{{\textrm{TM}}}(t)$ is actually a cyclostationary process, thus making $\mathcal{W}(t)$ cyclostationary as well. However, without loss of generality, in this paper we consider their stationary statistics by simply referring to the cyclic auto-correlation at null cyclic frequency~\cite{Pa91}.} to be a white process with power spectral density equal to $N_0$.   
Since $x_{\textrm{TM}}(t)$ is SRRC filtered, $|x_{\textrm{TM}}(t)|^2 \neq 1$, and thus perfect cancellation cannot be achieved.
However, taking into account that $\mathbb{E}[|x_{\textrm{TM}}(t)|^2] = 1$, sub-optimal cancellation can be performed by averaging the samples of the telemetry signal's time-varying envelope, as shown in Section~\ref{sec:timing}. 

\begin{figure}
\centering
\includegraphics[width=0.6\columnwidth]{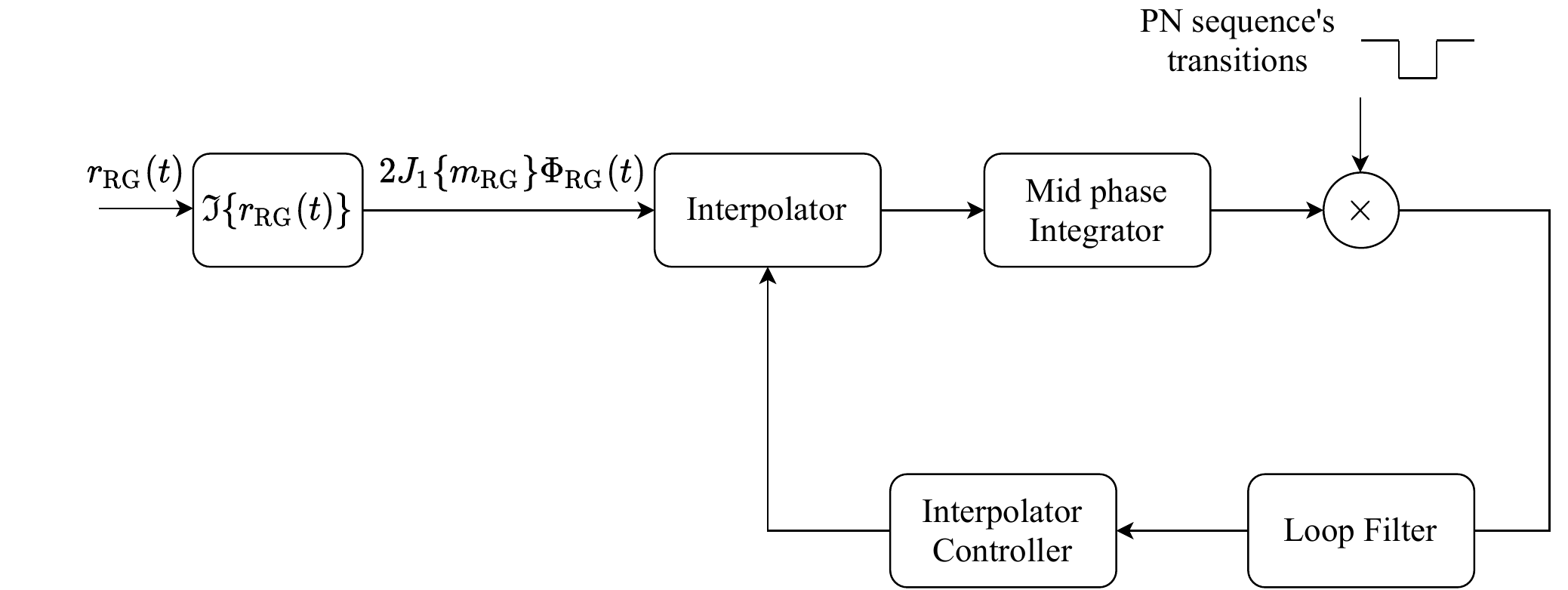}
\caption{Chip tracking loop block diagram.}
\label{fig:ctl}
\end{figure}

Once recovered, the ranging sequence is input to the Chip Tracking Loop (CTL), whose block diagram is depicted in Figure~\ref{fig:ctl}. Since the ranging chip sequence resembles a clock signal, we implement the CTL as a modified version of the Data Transition Tracking Loop (DTTL)~\cite{CCSDSPNG,JPL06}. In this case, the mid-phase integrator operates according to an Integrate and Dump approach, integrating the PN sequence over a time interval $T_c$ between two consecutive chips. Then, the mid-phase integrator output is multiplied by the PN transition sequence to adjust the sign of the error at the input of the loop filter.

Finally, the estimated CTL timing $\hat{\tau}_{\textrm{RG}}$ is fed back to the local PN generator in Figure~\ref{fig:Rx}, thus closing the receiver loop.  

\section{Ranging Timing Jitter} \label{sec:timing}

In this section we analyze the timing jitter of the receiver scheme presented in Section~\ref{sec:PSK/APSK+ranging}.

We define $\mathcal{P}(t) = |x_{\textrm{TM}}(t)|^2 - 1$ as a zero-mean signal with variance $\sigma^2_P = \mathbb{E}[|x_{\textrm{TM}}(t)|^2]$, which depends on the adopted telemetry modulation and is equal to the values reported in Table~\ref{tab:powJitt}.

\begin{table}
\begin{center}
\vspace{0.5cm}
\caption{Telemetry power variance $\sigma^2_P$ for different PSK/APSK modulations and roll-off factors.} 
\label{tab:powJitt}
\begin{tabular}{|c |c| c| c| c|c|}
\hline
Roll-off & QPSK & 8-PSK &16-APSK & 32-APSK & 64-APSK\\ \hline
0.2 & 0.240 & 0.243 & 0.435 & 0.560 &0.483\\
0.25 & 0.224 & 0.226  & 0.426 & 0.554 & 0.474 \\
0.3 & 0.209 & 0.211 & 0.418 & 0.551 & 0.467\\
0.35 & 0.197 & 0.199 & 0.412 & 0.549 & 0.462\\ \hline
\end{tabular}
\end{center}
\end{table}

Thus, Equation~\eqref{eq:PNtm_canc} can be re-written as

\begin{equation}
	y(t) x^*_{\textrm{TM}}(t) = r_{\textrm{RG}}(t-\tau_{\textrm{RG}}) + \mathcal{W}(t) +  r_{\textrm{RG}}(t-\tau_{\textrm{RG}}) \mathcal{P}(t) \,, \label{eq:in_CTL}
\end{equation}
where the last term can be understood as additional noise arising from the non-constant envelope of the considered modulation formats. To assess its impact, we analyze the output of the CTL. As Figure~\ref{fig:ctl} shows, the CTL extracts the imaginary component of the ranging signal, which reads

\begin{equation}
\alpha \Phi_{\textrm{RG}}(t-\tau_{\textrm{RG}}) + \mathcal{N}(t) + \mathcal{D}(t) \,,
\label{eq:ctl_im}
\end{equation}
where $\mathcal{N}(t)$ is the imaginary component of the Gaussian noise, $\mathcal{D}(t) = \alpha \Phi_{\textrm{RG}}(t-\tau_{\textrm{RG}})  \mathcal{P}(t)$, and\footnote{Considering sine-shaped chip pulses, applying the Jacobi-Anger expansion we have $r_{\textrm{RG}}(t) \approx J_0(m_{\textrm{RG}}) + j 2 J_1(m_{\textrm{RG}}) \Phi_{\textrm{RG}}(t)$~\cite{CCSDSGMSKPN_G}.}  $\alpha \approx 2 J_1(m_{\textrm{RG}})$.

Therefore, we define $n \triangleq \int_{0}^{T_c}{\mathcal{N}(t) \; \textrm{d}t}$ and $\zeta \triangleq \int_{0}^{T_c}{\mathcal{D}(t)} \;\textrm{d}t$, i.e, the output of the mid-phase integrator when $\mathcal{N}(t)$ and $\mathcal{D}(t)$ are input, respectively. While it is easy to see that $\mathbb{E}[|n|^2] = T_c N_0$, a closed-form solution for $\mathbb{E}[|\zeta|^2]$ is not easy to obtain but we derive its upper-bound. In fact, by considering that the auto-correlation of $\mathcal{D}(t)$ is always lower than $\mathbb{E}[|\mathcal{D}(t)|^2]$
\begin{equation}
\mathbb{E}[|\zeta|^2] \leq  \int_{0}^{T_c}{ \int_{0}^{T_c}{\mathbb{E}[|\mathcal{D}(t)|^2] \;\textrm{d}t_1 \textrm{d}t_2}} = P \sigma^2_P T_c^2\,,\ \label{eq:autocorrelation}
\end{equation}
with $P = \mathbb{E}[|\alpha \Phi_{\textrm{RG}}(t)|^2]$, i.e., the useful power of the received ranging signal at the input of the CTL.

\begin{figure}
\centering
	\includegraphics[width=0.6\columnwidth]{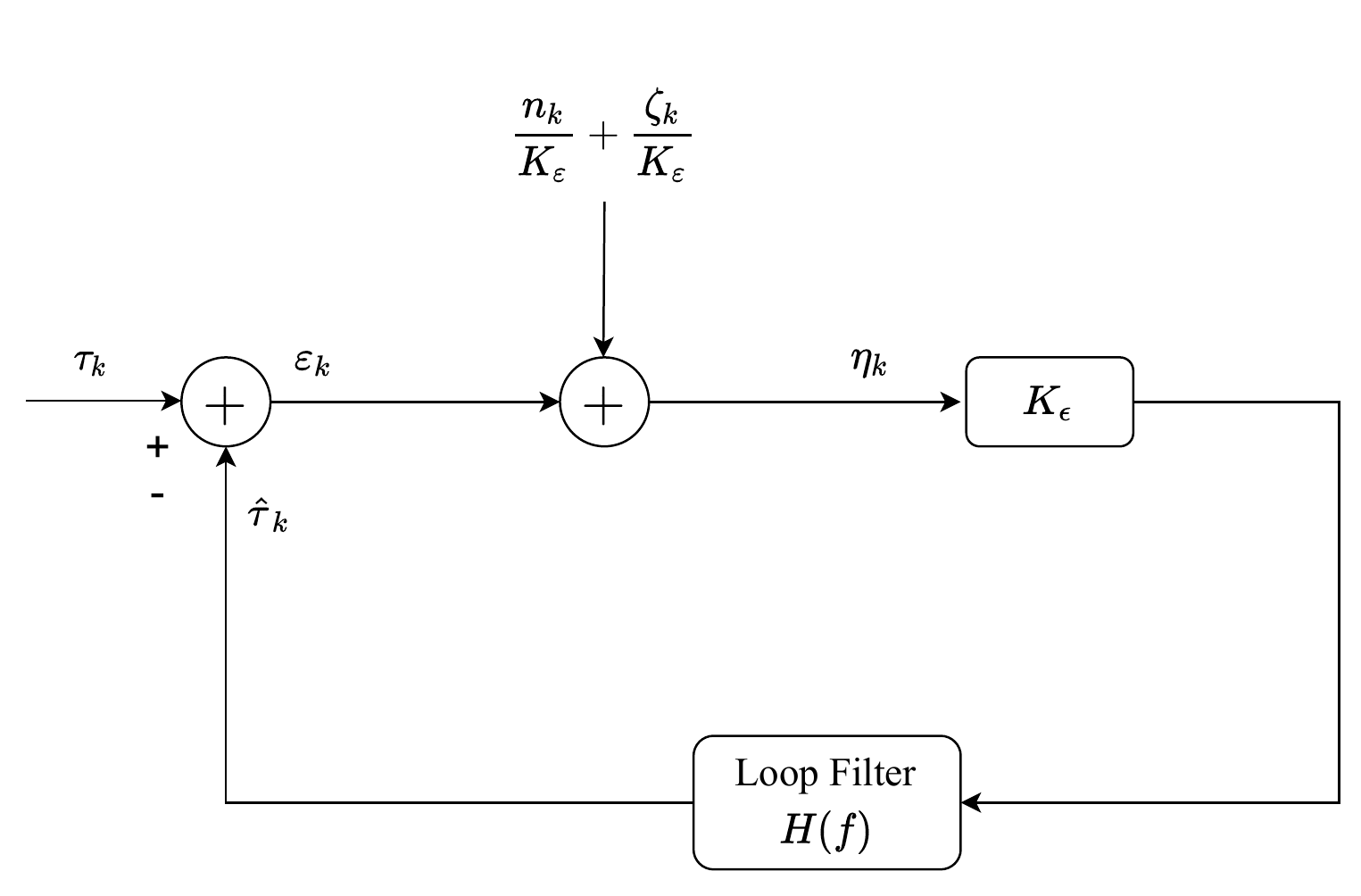}
	\caption{Linearized CTL model.}
	\label{fig:CTL-lin}
\end{figure}

We assume $\zeta$ to be a white process. Thus, we derive a linearized model of the CTL, as reported in Figure~\ref{fig:CTL-lin}, in which both the discretized noise terms  at the $k$-th time ($n_k$ and $\zeta_k$) are added to the phase detector output $\varepsilon _k$. 
The loop error is defined as
\begin{equation}
\eta_k = \varepsilon_k + \frac{n_k}{K_{\varepsilon}} +  \frac{\zeta_k}{K_{\varepsilon}} \,,
\end{equation}
where $K_{\varepsilon} = 2 \sqrt{2P}$ is the mid-phase integrator gain~\cite{CCSDSPNG}. \\ Considering that $n_k$ and $\zeta_k$ are uncorrelated, the loop error variance is $\sigma^2_{\eta} = (\mathbb{E}[|n_k|^2]+\mathbb{E}[|\zeta_k|^2])/K^2_{\varepsilon}$. 
Thus, by using Equation~\eqref{eq:autocorrelation},  and filtering the loop error $\eta_k$ through the Loop Filter with transfer function $H(f)$, the ranging timing variance is upper-bounded by
\begin{equation}
\sigma^2_{\tau} \leq T_c \int_{0}^{1/T_c}{\left|\frac{ K_{\varepsilon} H(f)}{1 + K_{\varepsilon} H(f)}\right|^2 \cdot T_c\left(\frac{N_0}{8 P} + \frac{\sigma^2_PT_c}{8} \right) df}\,.
\label{eq:autocorr}
\end{equation}
Recalling the definition of the noise bandwidth $B_L$~\cite{gardner2005} from Equation~\eqref{eq:autocorr} we obtain
\begin{equation}
{\sigma}^2_{\tau} \leq B_LT_c^2  \left( \frac{N_0}{8P} + \frac{ \sigma^2_PT_c}{8}\right)\,.
\label{eq:upperbound}
\end{equation}
It is interesting to observe that the first noise RHS term in Equation~\eqref{eq:upperbound} coincides with the theoretical PN ranging jitter~\cite{CCSDSPNG}, and is reduced as $P/N_0 B_L$ grows. However, due to the modulation's non-constant envelope, the second RHS term appears. As it is directly proportional to the product $\sigma^2_P B_L$, the only way to reduce this noise term is by narrowing the noise loop bandwidth $B_L$, i.e., by averaging the samples of the varying envelope.

\section{Numerical Results}
We simulate the transmission of a SRRC filtered telemetry signal, with channel symbol rate $R = 1/T = 4.2$ Msymbol/s, together with a PN ranging signal of rate $R_c = 1/T_c = 3$ Mchip/s.

First, we analyze the occupied bandwidth in an AWGN channel (as 99-percent of the signal power): in absence of ranging, the telemetry signal has approximately 4.8 MHz of bandwidth. Instead, in the presence of the PN signal, the spectral occupancy changes as a function of the ranging modulation index $m_{\textrm{RG}}$ as shown in Table~\ref{tab:99band}. For comparison, we include the bandwidth of a GMSK modulated signal (with the same bitrate of a QPSK modulated signal). It can be seen that, even in presence of ranging, PSK/APSK modulations are more spectrally-efficient than GMSK. Thus, when subject to spectral regrowth due to power amplification (not considered in this paper), the two schemes will have similar bandwidths.


\begin{table} 
\begin{center}
\vspace{0.5cm}
\caption{Occupied bandwidth for PSK/APSK (SRRC filtered with roll-off 0.35) and GMSK modulated telemetry, over the AWGN channel, as a function of the ranging modulation index $m_{\textrm{RG}}$.}
\label{tab:99band}
\begin{tabular}{|c| c| c|}
\hline
$m_{\textrm{RG}}$ & PSK/APSK & GMSK \\ \hline \hline
0.111 & 4.87 MHz & 7.22 MHz\\ 
0.222 & 5.04 MHz & 7.31 MHz\\ 
0.444 & 6.38 MHz & 7.72 MHz \\ 
0.666 & 7.31 MHz & 8.23 MHz \\ \hline
\end{tabular}
\end{center}
\end{table}

\begin{figure}
\centering
\begin{subfigure}{.5\textwidth}
  \centering
  \includegraphics[width=1.0\columnwidth]{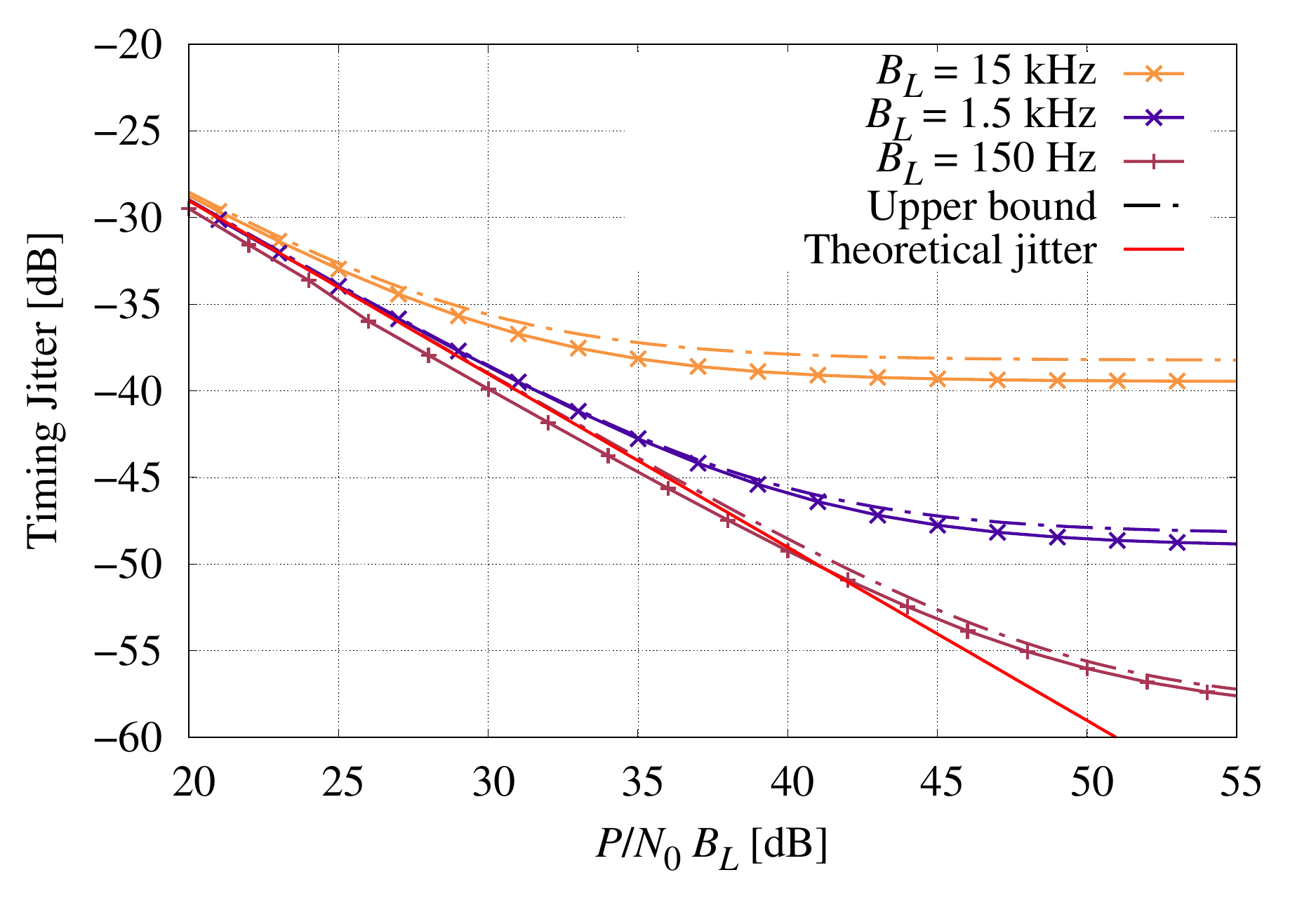}
  \caption{QPSK}
  \label{fig:QPSKupper}
\end{subfigure}%
\begin{subfigure}{.5\textwidth}
  \centering
  \includegraphics[width=1.0\columnwidth]{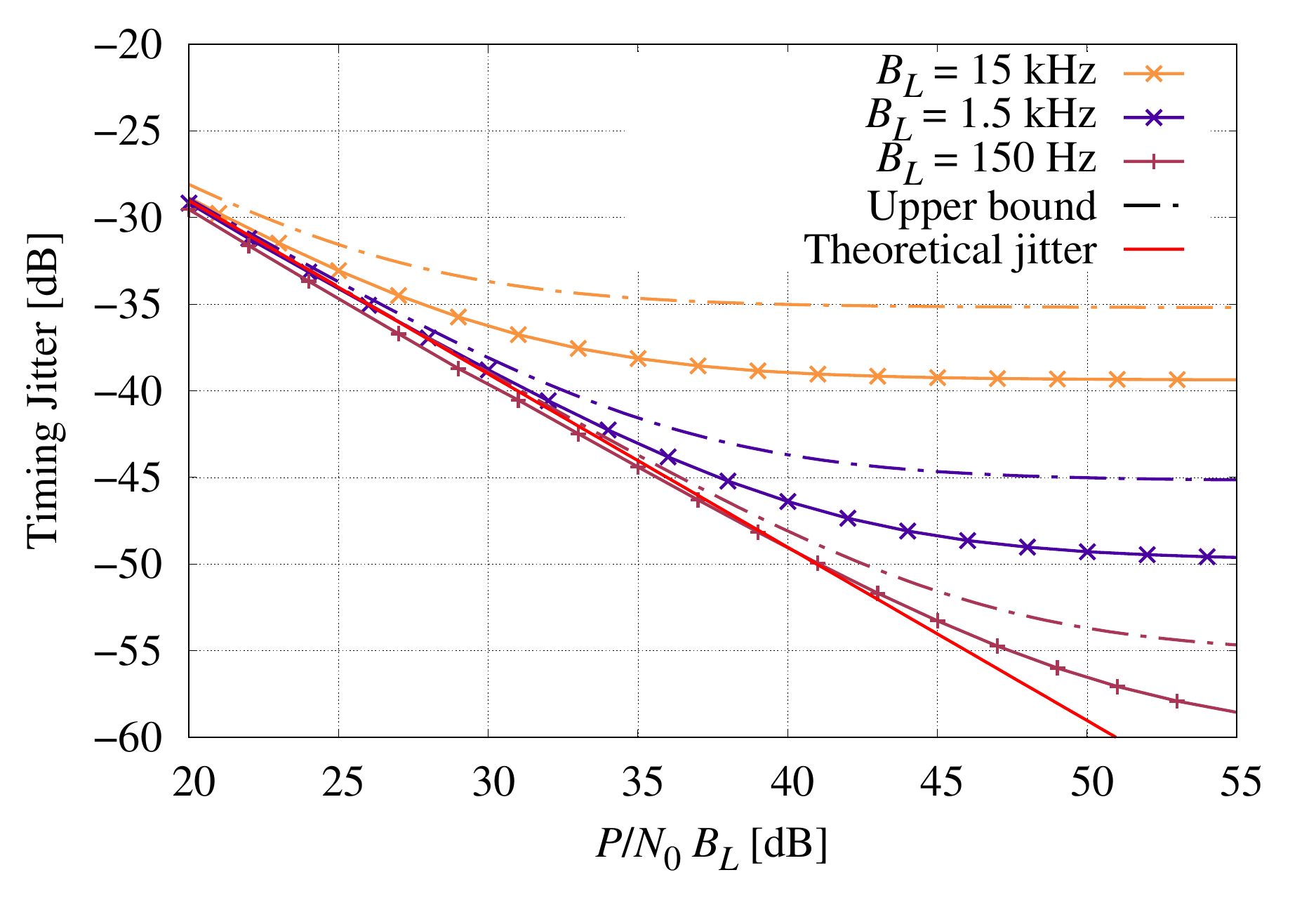}
  \caption{64-APSK }
  \label{fig:64upper}
\end{subfigure}
\caption{Impact of SRRC-filtered modulations on timing jitter in a genie-aided receiver (roll-off = 0.2).}
\label{fig:genieAidedresults}
\end{figure}

Next, we analyze the ranging jitter in the ideal condition of perfectly demodulated telemetry symbols. 
To do so, we implement and simulate a genie-aided receiver that is provided with the correct sequence of symbols $\{a_k\}$. We evaluate the timing jitter $\sigma^2_{\tau}$, normalized to $T_c^2$, as a function of the signal-to-noise ratio $P/N_0B_L$ and for different values of $B_L$. Figure~\ref{fig:genieAidedresults} shows the simulation results for SRRC filtered (roll-off = 0.2) QPSK (Figure~\ref{fig:QPSKupper}) and 64-APSK (Figure~\ref{fig:64upper}) modulations.
For comparison, the upper-bound of Equation~\eqref{eq:upperbound} is shown together with the theoretical PN ranging jitter derived in~\cite{CCSDSPNG}. As predicted, the jitter has a floor that decreases with $B_L$, and thus the performance gets closer to the theoretical jitter. In particular, a value $B_L=150$ Hz provides almost ideal performance, without limiting the timing synchronization dynamic. Additionally, Figure~\ref{fig:genieAidedresults} shows that as the modulation order decreases, the upper-bound is tighter to the actual jitter.

\begin{figure}
\centering
\begin{subfigure}{.5\textwidth}
  \centering
  \includegraphics[width=1.0\columnwidth]{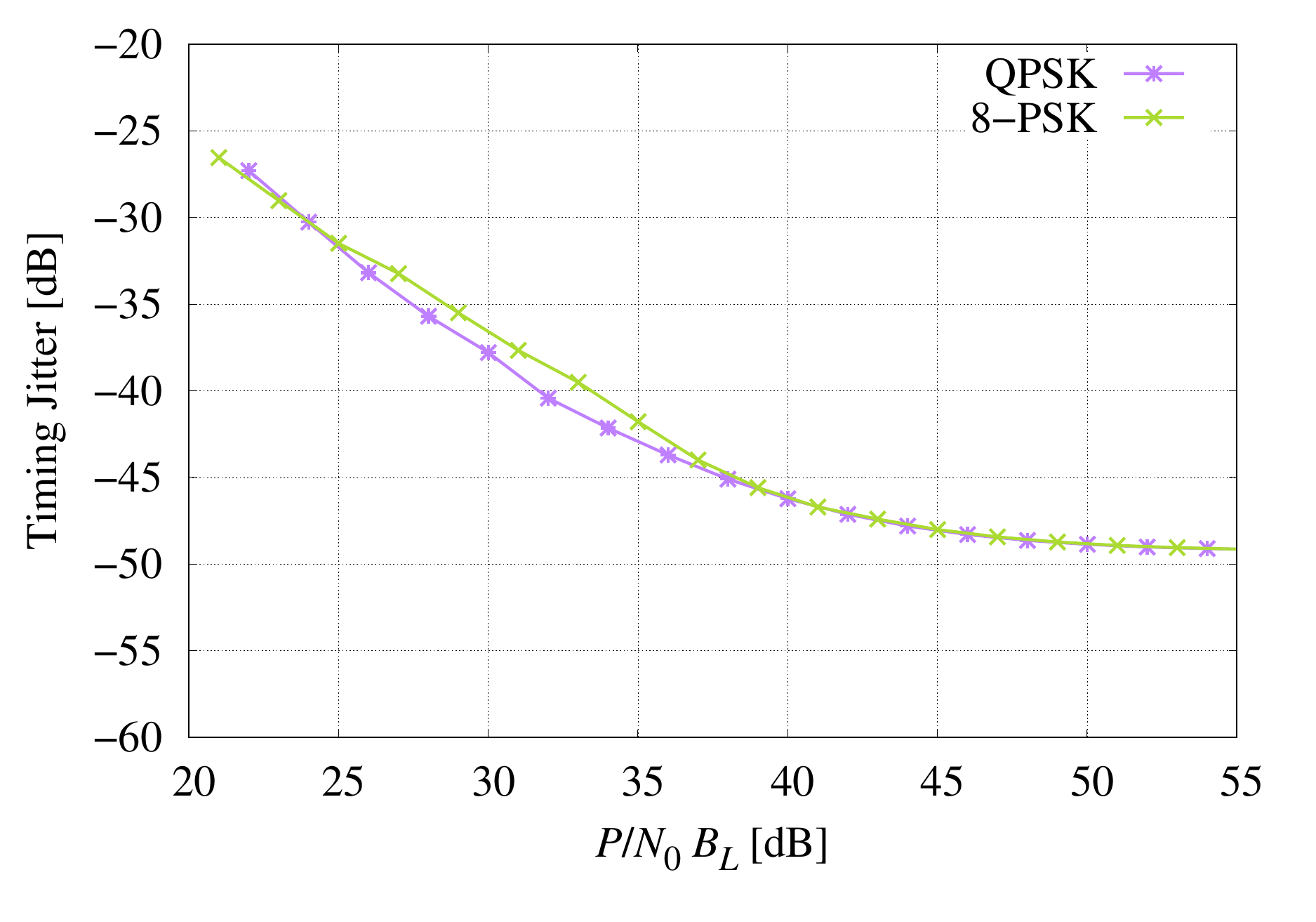}
  \caption{Ranging timing jitter}
  \label{fig:PSKjitter}
\end{subfigure}%
\begin{subfigure}{.5\textwidth}
  \centering
  \includegraphics[width=1.0\columnwidth]{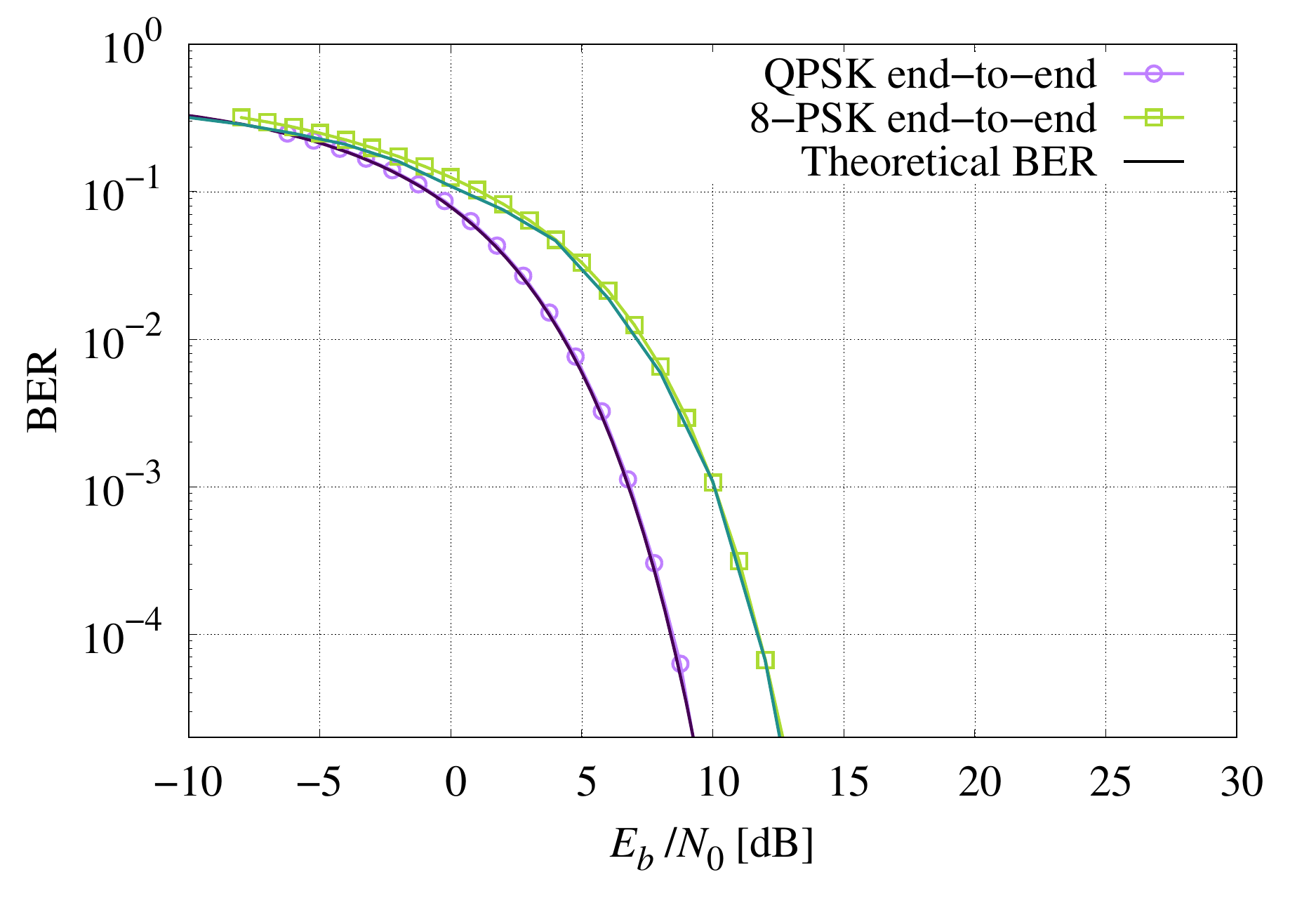}
  \caption{Telemetry BER}
  \label{fig:PSKber}
\end{subfigure}
\caption{End-to-end ranging and telemetry performance for QPSK and 8-PSK modulations ($B_L = 1.5$ kHz).}
\label{fig:PSKresults}
\end{figure}

\begin{figure}
\centering
\begin{subfigure}{.5\textwidth}
  \centering
  \includegraphics[width=1.0\columnwidth]{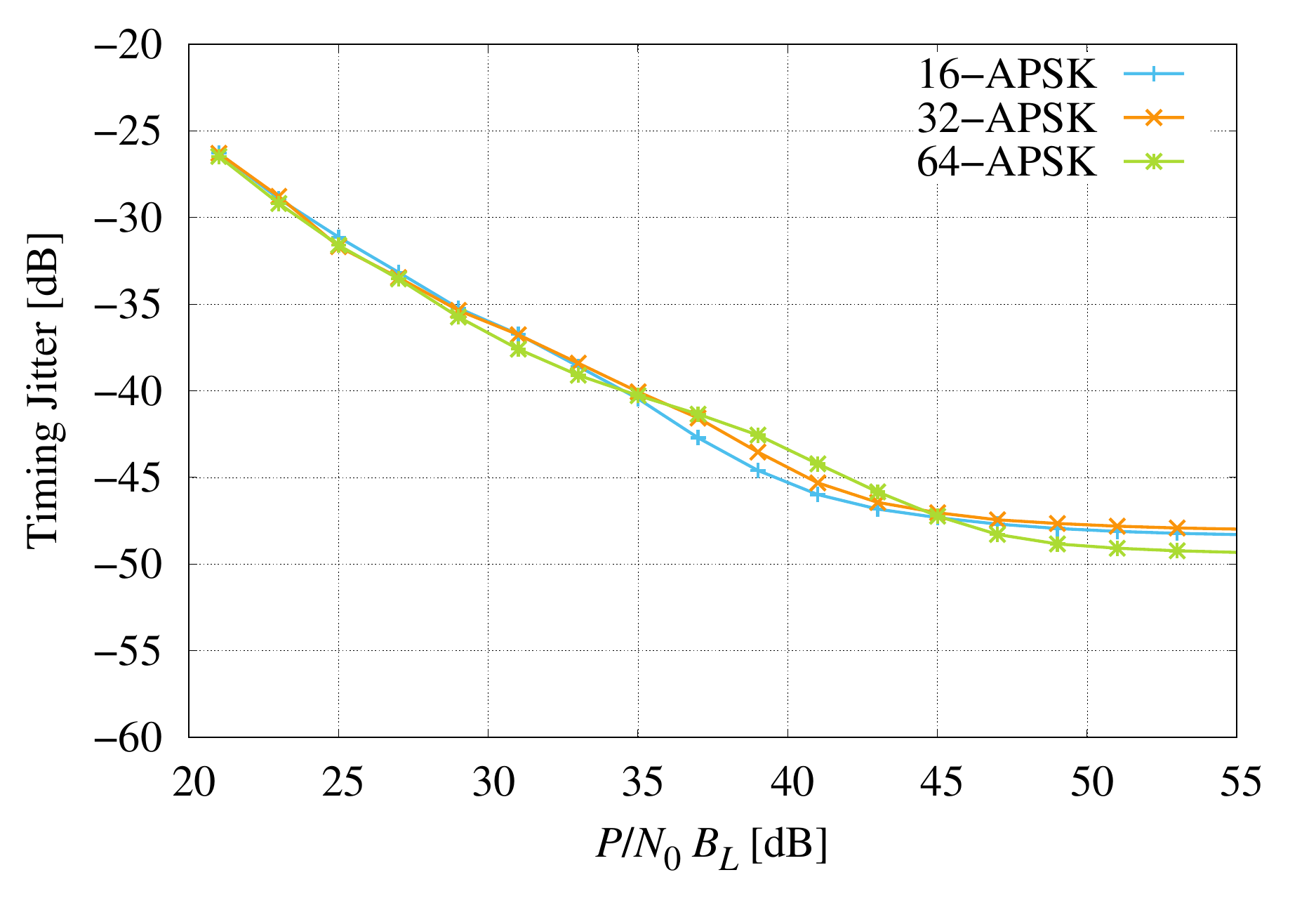}
  \caption{Ranging Timing Jitter}
  \label{fig:APSKjitter}
\end{subfigure}%
\begin{subfigure}{.5\textwidth}
  \centering
  \includegraphics[width=1.0\columnwidth]{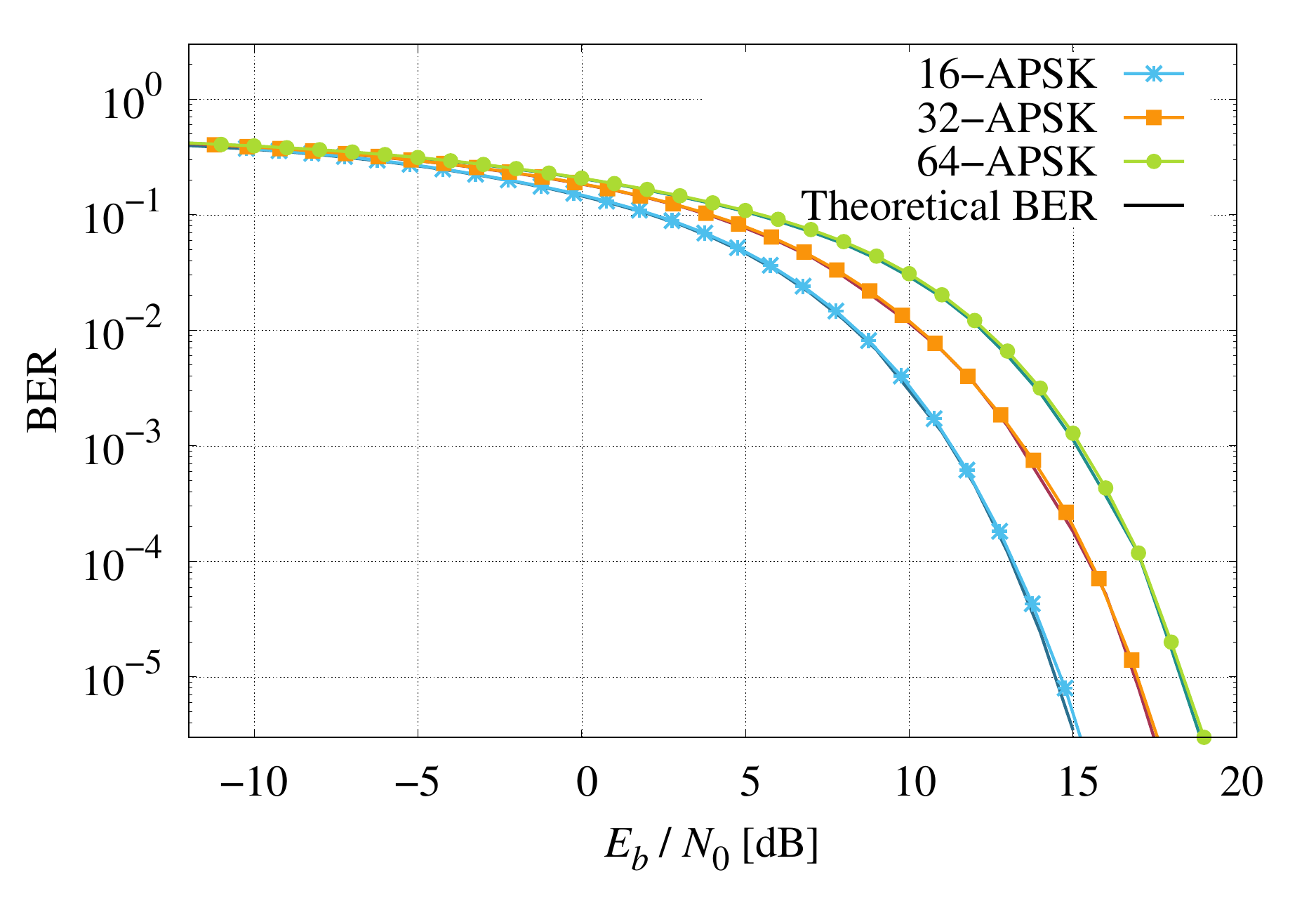}
  \caption{Telemetry BER}
  \label{fig:APSKber}
\end{subfigure}
\caption{End-to-end ranging and telemetry performance for 16-APSK, 32-APSK, and 64-APSK modulations ($B_L = 1.5$ kHz).}
\label{fig:APSKresults}
\end{figure}

Finally, we evaluate the end-to-end performance of the receiver scheme of Figure~\ref{fig:Rx} from the telemetry and ranging points of view, simulating the complete communication chain. Figure~\ref{fig:PSKresults} and \ref{fig:APSKresults} present the ranging timing jitter and the telemetry's BER as a function of $P/N_0 B_L$ and $E_b/N_0$ (being $E_b$ the energy per bit), respectively, for all SRRC-filtered PSK/APSK modulations (roll-off = 0.2).
For the two modulation sets the same trend is observed: despite an initial deviation (more pronounced as the modulation order grows), the simulated ranging jitter curves (Figure~\ref{fig:PSKjitter} and~\ref{fig:APSKjitter}) coincide with the genie-aided curve, which is free of telemetry losses.
Although the telemetry and ranging performances are inherently intertwined, the worsened jitter performance for low values of $P/N_0 B_L$, with respect to ideal cancellation, is not reflected in the BER. In fact, the end-to-end BER is comparable to the modulations' theoretical performance in an AWGN channel in the absence of ranging. (Figure~\ref{fig:PSKber} and~\ref{fig:APSKber}).
	
\section{Conclusions}
In this paper, we investigated the possibility of combining the PN ranging with high-order, spectrally-efficient PSK/APKS modulations. At the receiver side, we designed, implemented and simulated a receiver scheme for simultaneous demodulation of high-order telemetry symbols and tracking of the received ranging sequence for orbit determination. The obtained results show that it is possible to achieve ranging performance comparable to current state-of-the-art receiver architecture by simply lowering the noise loop bandwidth $B_L$. Furthermore, the introduction of such modulations for combined telemetry and ranging more than doubles/triples the current data rate bound of 8.7 Mbps.

\section*{Disclaimer}
	The view expressed herein can in no way be taken to reflect the official opinion of the European Space Agency.

\bibliographystyle{ieeetran}
\end{document}